\documentclass[
reprint,
superscriptaddress,
amsmath,
amssymb,
aps,
prx,
]{revtex4-2}
\usepackage{graphicx}
\usepackage{dcolumn}
\usepackage{bbm}
\usepackage{csquotes}
\usepackage[title]{appendix}
\usepackage[normalem]{ulem}
\usepackage{comment}
\usepackage{bm}
\usepackage{physics}
\usepackage{subfigure}
\usepackage[colorlinks,citecolor=blue]{hyperref}
\usepackage[capitalize]{cleveref}
\usepackage{comment}
\usepackage{todonotes}
\usepackage{fvextra}
\input{highlighted-code/pygments-style.tex}
\usepackage{xcolor}
\usepackage{standalone}

\usepackage{booktabs}
\usepackage{mathtools}
\usepackage{orcidlink}

\newcommand{\id}{\mathbb{I}}

\newcommand*{\xarg}{\ensuremath{\vb{x}}}

\newcommand{\Pf}{\operatorname{Pf}}

\usepackage{xspace}
\newcommand{\ggpeps}{\texttt{ggpeps}\xspace}
\newcommand{\numpy}{\texttt{numpy}\xspace}
\newcommand{\jax}{\texttt{jax}\xspace}
\newcommand{\scipy}{\texttt{scipy}\xspace}
\renewcommand{\vec}[1]{\ensuremath{\mathbf{#1}}}
\renewcommand{\norm}[1]{\left\lVert#1\right\rVert}

\usepackage{tikzbubbles} 
\usetikzlibrary{calc}

\begin{document}

\title{ggpeps: A package for Gauged Gaussian PEPS}

\newcommand{\aqa}{$\langle aQa ^L\rangle $ Applied Quantum Algorithms, Universiteit Leiden}
\newcommand{\lorentz}{Instituut-Lorentz, Universiteit Leiden, Niels Bohrweg 2, 2333 CA Leiden, Netherlands}
\newcommand{\ulm}{Institute for Complex Quantum Systems, Ulm University, 89069 Ulm, Germany}
\newcommand{\iqst}{Center for Integrated Quantum Science and Technology (IQST), Ulm-Stuttgart, Germany}

\author{Ariel Kelman\orcidlink{0000-0002-7710-6538}}
\affiliation{Racah Institute of Physics, The Hebrew University of Jerusalem, Givat Ram, Jerusalem 91904, Israel}
\author{Itay Gomelski\orcidlink{0009-0002-4811-9660}}
\affiliation{School of Physics and Astronomy, Tel Aviv University, Tel Aviv 6997801, Israel}
\author{Jonathan Elyovich\orcidlink{0009-0006-0123-4505}}
\affiliation{School of Physics and Astronomy, Tel Aviv University, Tel Aviv 6997801, Israel}
\author{Erez Zohar\orcidlink{0000-0001-6993-6569}}
\affiliation{School of Physics and Astronomy, Tel Aviv University, Tel Aviv 6997801, Israel}
\author{Patrick Emonts\orcidlink{0000-0002-7274-4071}}
\affiliation{\ulm}
\affiliation{\iqst}
 
\date{\today}

\begin{abstract}
Lattice gauge theories are a low-energy regularization of quantum field theories. 
They are the basis for the development of numerical algorithms that provide insight into the non-perturbative regime of quantum field theories which is difficult to achieve through other methods.
We present a python package, \ggpeps, implementing gauged Gaussian projected entangled pair states (GGPEPS), specialized to study lattice gauge theories. 
We give a an overview of the considered systems, describe the GGPEPS ansatz and provide a usage guide for the \ggpeps package.
\end{abstract}
\maketitle

\section{Introduction}
Lattice gauge theories are an important tool in numerically studying a wide class of systems that are of interest in high energy and condensed matter systems~\cite{wilson_confinement_1974}.
One standard technique involves calculating expectation values by evaluating the path integral of interest, usually with Monte Carlo techniques~\cite{creutz_monte_1979, creutz_monte_1983,Banuls-ReviewNovelMethods-2020,alexandru_complex_2020}.
This approach, however, runs into challenges, especially the sign problem~\cite{troyer_computational_2005}, in which the quantity intended for use as a probability distribution takes on complex or negative values.
Another issue arises for studying time evolution, as the Wick rotation necessary for applying Monte Carlo redefines the time coordinate, thereby making real-time evolution difficult.

A Hamiltonian approach to lattice gauge theories solves these problems~\cite{kogut_hamiltonian_1975}, but comes with other challenges.
Instead of discretizing time and space, time is kept continuous.
Given the Hamiltonian framework, we cannot compute with the action, i.e. a scalar, but we have to find methods to efficiently encode or approximate states to estimate observables of interest.
This can be accomplished in different ways: either by simulating the many-body systems on a classical computer or by encoding the system into quantum hardware~\cite{dalmonte_lattice_2016,Banuls-SimulatingLatticeGauge-2020,zohar_quantum_2021,di_meglio_quantum_2024,klco_standard_2022}. 
Here, we focus on variational classical techniques, specifically tensor networks.

Tensor networks are a class of states designed for many-body physics, which captures entanglement properties 
and has been very successful in many applications~\cite{cirac_matrix_2021}.
Considerable effort has been devoted to the numerical implementation of tensor network algorithms, alongside the development of efficient algorithms to optimize them.
Algorithms like the Density Matrix Renormalization Group (DMRG), which finds ground states using Matrix Product States (MPS)~\cite{White-DensityMatrixFormulation-1992, White-DensitymatrixAlgorithmsQuantum-1993}, have proven tremendously influential. 
Tensor networks have been employed to investigate the time evolution of quantum systems~\cite{Zwolak-MixedStateDynamicsOneDimensional-2004, Daley-TimedependentDensitymatrixRenormalizationgroup-2004}, thermal states~\cite{Verstraete-MatrixProductDensity-2004}, and other applications 
~\cite{bou-comas_quantics_2025,niedermeier_solving_2025,gomez-lozada_simulating_2025, Bacciconi-PullingStringsReal-2026}.

Given their success, tensor networks have also been applied to lattice gauge theory systems~\cite{haegeman_gauging_2015,silvi_lattice_2014,zohar_fermionic_2015,tagliacozzo_tensor_2014,kuhn_quantum_2014,Banuls-SimulatingLatticeGauge-2020,magnifico_tensor_2025,wu_accurate_2025}.
Of particular interest here are demonstrations of tensor networks applied to systems suffering from the sign problem~\cite{Banuls-OvercomingMonteCarlo-2017}, as well as their application to simulating high dimensional systems, frequently employing tensor network variants, including tree tensor networks~\cite{Felser-TwoDimensionalQuantumLinkLattice-2020,magnifico_lattice_2021} and infinite projected entangled pair states (PEPS)~\cite{Robaina-Simulating-2021}. 
The goal of this paper is to introduce a python package, \ggpeps, built around a specific tensor network ansatz, gauged Gaussian PEPS (GGPEPS), designed for use with lattice gauge theories in a Hamiltonian (rather than action-based) framework.

The \ggpeps package provides a vectorized and GPU-aware Python implementation of GGPEPS, presented comprehensively in Ref.~\cite{kelman_gauged_2024}. 
A more detailed review and references are provided in section~\ref{sec:system}; here we note the main ideas: we build a tensor network state that avoids the need for explicit contraction, and instead express the state as a superposition of many Gaussian states, each of which can be captured by a covariance matrix (matrix of second order moments). 
Utilizing a variational Monte Carlo procedure (see section~\ref{sec:mc}), calculations can be efficiently performed on the full state.
One common operation, demonstrated in Refs.~\cite{emonts_finding_2023, kelman_projected_2026}, is a ground state search.
Here, we introduce the code, provide usage examples, and note some important details relating to the implementation. 
The code can be found at~\cite{Kelman-GGPEPSGithubSource-2026}.
The \ggpeps package does not substitute other tensor network packages like \texttt{itensor}, \texttt{quimb}, \texttt{tensorkit}, \texttt{tenpy} or other frameworks~\cite{itensor,devos_tensorkitjl_2026,gray_quimb_2018,hauschild_efficient_2018}.
It rather supplements it by exploring a specialized set of states that has gauge invariance strictly built into the construction.

The rest of the paper is structured as follows: in \Cref{sec:system} we describe the systems for which the GGPEPS ansatz is designed, as well as the ansatz itself, including implementation details.
In \Cref{sec:mc} we discuss the Monte Carlo sampling method our approach utilizes, and an important detail -- ``local updates'' -- that greatly improves the efficiency of calculation.
In \Cref{sec:code} we present the code itself, together with a usage guide for some basic examples, and then point out some important optimizations as well as future directions.

\section{Definition of the System}
\label{sec:system}

\subsection{System}

\begin{figure}[h]
	\includegraphics[width=0.7\linewidth]{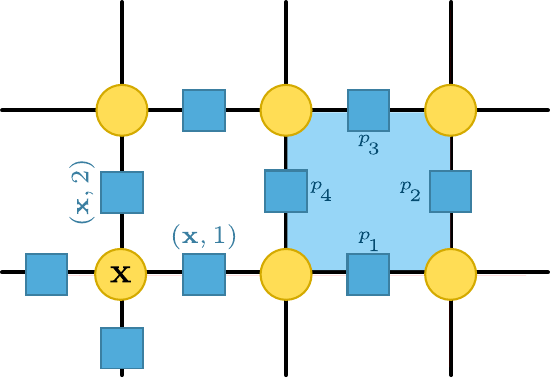}
	\caption{A diagram of the 2D lattice. Matter is shown on lattice sites in yellow and gauge fields on links in blue. The blue box shows the labeling convention for a plaquette.}
    \label{fig:lattice}
\end{figure}

We consider a lattice gauge theory on a square (or cubic) lattice with periodic boundary conditions. 
Such a system has gauge fields on the lattice links (elements of the gauge group under consideration), and fermionic matter on the vertices, as shown in \cref{fig:lattice}. 

To such a system corresponds a gauge group $G$ and a Hamiltonian, composed of terms which depend on the gauge fields and matter content of the system.
The Hamiltonians we consider have the form 
\begin{equation} \begin{aligned}
    H &= g_{E} H_{E} + g_{M} H_{M} + g_{I} H_{I} + g_{m} H_{m} + H_\mu
\end{aligned} \end{equation}
where the first two terms, $H_E$ and $H_B$, depend only on the gauge fields (denoted electric and magnetic respectively), the interaction term $H_I$ depends on both the gauge fields and matter, and the last two terms, $H_m$ and $H_\mu$, depend only on the matter and represent the mass and chemical potential energy.

The system, including the Hamiltonian, will have several symmetries, which we make use of in the next section to arrive at an ansatz that expresses states with these symmetries in an efficient manner.
The symmetries may include translation (perhaps by more than a single site), rotation (by multiples of $\pi/2$ since we work on square lattices), a $U(1)$ fermionic number conservation symmetry (if desired), and the local gauge symmetry that defines the theory --- puts the ``G'' in ``LGT''. This local symmetry adds constraints at each lattice site. 
These are generally hard to encode, but the GGPEPS ansatz obeys them by construction.

The general structure of the ansatz, as well as the code, is general enough to support any gauge group (i.e. a finite or compact Lie group). 
Currently, the $\mathbb{Z}_2$ is the most well-developed, though the code already contains support for Dihedral groups without matter. 
In the remainder of this paper, we focus on the $\mathbb{Z}_2$ example, but mention the necessary changes required for other groups.

\subsection{Ansatz}

The key ingredient for a variational algorithm is the family of variational states that can be expressed.
We aim to construct a state that
\begin{enumerate}
    \item enables efficient computation;
    \item obeys the Gauss law and other symmetries by construction;
    \item obeys an entanglement area (boundary) law, as expected for the low energy states of local gapped Hamiltonians~\cite{hastings_area_2007,eisert_colloquium_2010}.
\end{enumerate}
To do so, we construct a tensor network state based on a projected entangled pair state~\cite{verstraete_renormalization_2004}. 
We add virtual degrees of freedom on each link around each site, and couple them to the physical matter and gauge fields as we now briefly describe.

The first of the requirements is achieved by constructing the ansatz state out of Gaussian components which allows the use of a covariance matrix formulation, since Gaussian states are entirely captured by their covariance matrix~\cite{bravyi_lagrangian_2005}.
The second and third are achieved through careful construction of the state --- some important details are noted below, but we refer to earlier more detailed work for this aspect of the construction~\cite{kelman_gauged_2024}.
In that work, it was shown that the state, with parameters $\alpha$,
\begin{equation} \begin{aligned}
    \label{eq:ansatz-integral-form}
    \ket{\Psi_\alpha}
        &= \int \mathcal{D} \mathcal{G} \ \ket{\psi_\alpha(\mathcal{G})} \ket{\mathcal{G}}, \\
\end{aligned} \end{equation}
where $\ket{\mathcal{G}}=\prod_l\ket{g_l}$ defines the gauge field configuration on the lattice with edges labeled by assigning a group element $g \in G$ to each link, while $\ket{\psi_\alpha(\mathcal{G})}$ defines the state of physical matter on the lattice sites.
It is $\ket{\psi_\alpha(\mathcal{G})}$ which is Gaussian (not the full state $\ket{\psi}$), and can be capture by its covariance matrix. 

We build this $\ket{\psi_\alpha(\mathcal{G})}$ as a tensor network, by introducing new fermionic Fock Hilbert spaces for each link emanating from a site. 
On a 2-dimensional lattice, we label these new ``virtual'' modes $r^\dagger,u^\dagger,l^\dagger,d^\dagger$, which are the modes \textit{right, left, up, down} of a site (we allow multiple modes --- copies --- per site and link, in which case we index them appropriately).
These operators will couple to each other, to the physical mode $\psi^\dagger$, and to the gauge fields, to build $\ket{\psi_\alpha(\mathcal{G})}$ as
\begin{equation} \begin{aligned}
    \label{eq:ansatz-for-gauge-config}
    \ket{\psi_\alpha(\mathcal{G})}
        &= \prod_{\text{layer}} \Big[ \bra{\Omega_v} \prod_l \omega(l) \prod_l U^\mathcal{G}(l) \prod_{\vec{x}} A_\alpha(\vec{x}) \ket{\Omega_v} \ket{\Omega_p}\Big], \\
\end{aligned} \end{equation}
where
\begin{equation} \begin{aligned}
    \label{eq:ansatz-components}
    A_\alpha(\vec{x}) &= \exp \Big( \mathcal{T}^\alpha_{ij}(\vec{x}) 
        a^\dagger_i(\vec{x}) a^\dagger_j(\vec{x}) \Big), \\
    U^\mathcal{G}(\vec{x}, k) &= \int dg \ket{g}\bra{g}_{\vec{x},k} \otimes \mathcal{U}_g \left(\vec{x},k\right), \\
    \omega(\vec{x}, k) &= \exp \Big( W^{(k)}_{ij}
    a^{\dagger}_i(\vec{x})
    a^{\dagger}_j(\vec{x} + \vec{e}_k)
    \Big).
\end{aligned} \end{equation}
Here, $\mathcal{U}_g$ implements the gauging operation for a fixed gauge group element $g$ on the virtual modes of the given link, guaranteeing the local gauge symmetry (for more details, see~\cite{kelman_gauged_2024}).
The Fock vacua of the virtual and the physical modes are denoted by $\ket{\Omega_v}, \ket{\Omega_p}$, respectively.

The parameters $\alpha$ of the state are entirely contained in the matrices $\mathcal{T}^\alpha(\vec{x})$. 
The sparse matrix $W^{(k)}$ contains fixed values which must be compatible with the desired symmetries.
The product over layers in \cref{eq:ansatz-for-gauge-config}, explained more below, comes into play when it is possible to partition all of the modes such that the operators $A_\alpha, \mathcal{U}^\mathcal{G}, \omega$ (which are all indexed by the layer, though this is suppressed in the notation) only act on one part, and so the entire expression trivially factorizes. 
In what follows, we generally drop the explicit indication of the dependence (of operators, states) on the state parameters $\alpha$ to keep the notation clean.

Representing and manipulating this state in a form amenable to numerical calculations is the purpose of the \ggpeps package. 
In the remaining part of this section, we provide an overview of the entire pipeline from defining parameters of the state to calculating an observable. 

The most general $\mathcal{T}(\vec{x})$ matrix is simply a fully general complex valued $n \times n$ matrix, where $n$ is the number of physical and virtual degrees of freedom associated with a site. 
Several constraints serve to limit this quite significantly: fermionic antisymmetry, translation and rotation invariance, and (if desired) a $U(1)$ symmetry.
For the case of several copies of virtual modes, with a single physical fermionic flavor per site, the structure of $\mathcal{T}(\vec{x})$ was worked out in detail in~\cite{kelman_projected_2026}.

By way of example, we can assume that the form of the $\mathcal{T}$ matrix has been determined for a 1-copy ansatz with a single flavor of fermionic matter on each site. 
Each $\mathcal{T}(\vec{x})$ can be represented as a $5 \times 5$ matrix: 1 physical mode, and a virtual mode in each direction:
\begin{equation} \label{eq:t-mat-example}
    \mathcal{T} = \begin{pmatrix}
                0 & it_{i} & -t_{i} & -it_{i} & t_{i} \\
                - it_{i} & 0 & -z & -iy & -iz \\
                t_{i} & z & 0 & -iz & y \\
                it_{i} & iy & iz & 0 & z \\
                -t_{i} & iz & -y & -z & 0 \\
            \end{pmatrix}.
\end{equation}
For such a system, $y,z$ define the variational parameters of the ansatz.
The number of parameters is significantly reduced due to the symmetries imposed by the system.
The mode order is $\psi^\dagger, r^\dagger,u^\dagger,l^\dagger,d^\dagger$, which are as before the modes \textit{right, left, up, down} of each site.
Note that this form of the $\mathcal{T}$ matrix violates the $U(1)$ symmetry, since it is impossible to specify how the modes transform under $U(1)$ while guaranteeing that products do not change; doing so requires adding extra copies, as detailed in~\cite{kelman_projected_2026}.

The values of the parameters in the $\mathcal{T}$ matrix, together with a gauge configuration $G$ define the state $\ket{\psi_\alpha(G)}$.
In particular, they determine the covariance matrix of this state, from which all observables can be calculated.
Define $\Gamma_{\textnormal{v}}(\vec{x})$ as the covariance matrix of the state
\begin{equation} \begin{aligned}
    \ket{\varphi(\vec{x})}
        &= A(\vec{x}) \ket{\Omega_v} \ket{\Omega_p}. \\
\end{aligned} \end{equation}
It can be constructed directly from $\mathcal{T} = \mathcal{T}(\vec{x})$ (as shown in Appendix B of~\cite{emonts_finding_2023}), as
\begin{equation}
\label{eq:gamma_dirac}
  \Gamma^D(\vec{x}) = i\begin{pmatrix}
            -\mathcal{T}^- \mathcal{T} & \frac{1}{2} \mathcal{T}^- (1 + \mathcal{T}\bar{\mathcal{T}}) \\
            -\frac{1}{2} \mathcal{T}^- (1 + \bar{\mathcal{T}} \mathcal{T}) & \mathcal{T}^- \bar{\mathcal{T}}
        \end{pmatrix}
\end{equation}
where $\mathcal{T}^- = (1 - \mathcal{T}\bar{\mathcal{T}})^{-1}$ (and $\bar{\mathcal{T}}$ is the complex conjugate of $\mathcal{T}$).

Majorana covariance matrices are guaranteed to have only real entries, which both reduces their memory footprint and provides a useful check on the computation for testing purposes.
We therefore prefer to work with Majorana modes, defined in terms of the Dirac modes used until now as
\begin{equation} \begin{aligned}
\label{eq:dirac-majorana-modes}
    c^\dagger &= \frac{1}{2} (c^{(1)} + ic^{(2)})
    \quad \quad &
    c^{(1)} &= c + c^\dagger \\
    c &= \frac{1}{2} (c^{(1)} - ic^{(2)})
    \quad \quad &
    c^{(2)} &= i(c - c^\dagger). \\
\end{aligned} \end{equation}
This can be used to convert the Dirac covariance matrix $\Gamma^D(\vec{x})$ into the Majorana one $\Gamma^M(\vec{x})$.

For calculations, we wish to have the full covariance matrix $\Gamma_{\textnormal{v}}$ of
\begin{equation} \begin{aligned}
    \ket{\varphi}
        &= \prod_{\vec{x}} A(\vec{x}) \ket{\Omega_v} \ket{\Omega_p}. \\
\end{aligned} \end{equation}
which can be constructed as
\begin{equation} \begin{aligned}
\label{eq:expand_to_system}
    \Gamma_{\textnormal{v}} = \bigoplus_{\vec{x}} \Gamma(\vec{x}).
\end{aligned} \end{equation}
in the Dirac or Majorana version as desired.
It is necessary for some calculations, e.g. \cref{eq:physical-fermions-cov} below, to extract blocks of $\Gamma_{\textnormal{v}}$ corresponding to the correlations of physical modes among themselves, physical with virtual, or virtual modes among themselves; we denote these $A, B, D$, respectively.

Next we define the covariance matrix $\Gamma_\text{in}(\mathcal{G})$ of the state 
\begin{equation} \begin{aligned}
    \ket{\phi(\mathcal{G})}
        &= \prod_l (U^\mathcal{G})^\dagger(l) \prod_l w^\dagger(l) \ket{\Omega_v}. \\
\end{aligned} \end{equation}
To calculate $\Gamma_\text{in}(\mathcal{G})$, we first compute the covariance matrix $\Gamma'_\text{in}$ of the state 
\begin{equation} \begin{aligned}
    \ket{\phi'}
        &= \prod_l w^\dagger(l) \ket{\Omega_v}. \\
\end{aligned} \end{equation}
analytically using Majorana modes. 
These covariance matrices are fixed in the simulation. 
As in the case of the $A(\vec{x})$ operator, $\Gamma'_\text{in}(\vec{x})$ is first calculated per link, and then the result is combined into a system-wide covariance matrix. 
In all previous work with this ansatz, the projectors $\omega(l)$ have been translationally invariant (though differing between links in different lattice directions), and the code assumes this behaviour.

The gauging operation, i.e. calculating $\Gamma_\text{in}(\mathcal{G})$ from $\Gamma'_\text{in}$ is a mixing operation of the modes on each link $l = (\vec{x}, k)$ depending on the gauge field value on each link respectively. 
The exact form of this operation depends on the gauge group, and we refer the interested reader to past work~\cite{kelman_gauged_2024, kelman_projected_2026}. 

From these components, one can construct the fermionic covariance matrix of the state $\ket{\psi_\alpha (G)}$ of \cref{eq:ansatz-for-gauge-config} as
\begin{equation} \begin{aligned}
\label{eq:physical-fermions-cov}
    \Gamma(\mathcal{G})
        &= \frac{i}{2} \langle [c^{(i)}(\vec{x}), c^{(j)}(\vec{y})] \rangle \\
        &= A + B (D + \Gamma_{\textnormal{in}}(\mathcal{G}))^{-1} B^T.
\end{aligned} \end{equation}

In some cases, the $\mathcal{T}$ matrix will have a block diagonal structure. 
This can be motivated by targeting different copies of extra modes for different physical purposes --- e.g. some for handling pure-gauge physics, and others for handling interactions with each flavor of physical modes.
If the gauging and projection operations similarly do not couple between sets of extra modes, then all covariance matrices of the full system will be block diagonal. 
In such a case, it is computationally efficient to separate the calculations into ``layers''; that is, to stack each stage of the computation into an extra axis, one per layer, each of which only deals with a subset of the modes.
This greatly reduces the size of the matrices that must be constructed. 
In \cref{sec:code} the label \texttt{\_vec} appended to a variable name often indicates an array dimension corresponding to the number of layers. 

For each gauge field configuration and associated matter state $\ket{\psi(\mathcal{G})}$, the expectation value of any observable can be calculated as function of the field configuration together with the covariance matrix of the matter state.
To do so, we require the norm of the state $\ket{\psi_\alpha (\mathcal{G})}$, which is given by
\begin{equation} \begin{aligned}
\label{eq:norm}
    \norm{\ket{\psi_\alpha (\mathcal{G})}}^2
    &= \sqrt{\det\left(\frac{1 - \Gamma_\text{in}(\mathcal{G})D}{2}\right)} \\
    &= \sqrt{\det (D) \det\left(\frac{D^{-1} - \Gamma_\text{in}(\mathcal{G})}{2}\right)} 
\end{aligned} \end{equation}
where $D$ is the submatrix of $\Gamma_\text{v}$ containing the correlations of all virtual modes among themselves. 
In the case of multiple layers, the total norm is the product of the norm of all layers.
With this norm in hand, the expectation value of an observable $\mathcal{O}$ can always be put into the form
\begin{equation} \begin{aligned}
\label{eq:observable}
    \langle \mathcal{O} \rangle
    &= \int \mathcal{DG} \ F_\mathcal{O}(\mathcal{G}) p(\mathcal{G}),
\end{aligned} \end{equation}
where $p(\mathcal{G})$ represents a probability distribution and is proportional to the expression in ~\cref{eq:norm}. 
When evaluating by integral by summing all the terms (only possible for finite groups and small system sizes) the normalization factor is given by the sum of all norms over $\mathcal{G}$. 
As we discuss in the next section, for a Monte Carlo simulation we do not require the normalization factor.
The calculation of other observables --- that is, of the $F_\mathcal{O}(\mathcal{G})$ --- from the quantities discussed here is given in previous work, such as the electric energy in~\cite{emonts_finding_2023}, or the interaction and mass energy in~\cite{kelman_projected_2026}. 
Since the implementation of the electric energy is far from trivial, especially in an efficient manner than generalizes to different gauge groups, \cref{sec:electric_implementation} gives some details on that implementation.

\begin{figure}
  \centering
  \begin{tikzpicture}[scale=1.5]

    \node[draw,
        minimum width=2cm,
        minimum height=1cm, align=center] at (3.6,1.1) (block0c) {State parameters \\ \texttt{system.cfg.paramvec}};

    \node[draw,
        minimum width=2cm,
        minimum height=1cm,
        align=center] at (0.65,1.1) (block1a) 
        {$\Gamma'_\text{in}(\vec{x})$ \\\texttt{system.gamma\_gauge\_neutral\_vec}};

    \node[draw,
        minimum width=2cm,
        minimum height=1cm,
        align=center] at (0.3,0) (block1b) {Gauge configuration $\mathcal{G}$ \\ \texttt{system.gaugefieldvec}};


    \node[draw,
        minimum width=2cm,
        minimum height=1cm,
        align=center] at (0.8,-2.2) (block3a) 
        {$\Gamma_\text{in}$ \\ \texttt{system.gamma\_in\_sys\_vec}};

    \node[draw,
        minimum width=2cm,
        minimum height=1cm,
        align=center] at (3.6,0) (block1c) 
        {$\mathcal{T}(\vec{x})$ matrices \\ \texttt{system.tmat\_layervec\_sitevec}};

    \node[draw,
        minimum width=2cm,
        minimum height=1cm,
        align=center] 
        at (3.6,-1.1) (block2c) 
        {$\Gamma_\textnormal{v}(\vec{x})$ in Dirac modes \\
        \texttt{system.gamma\_dirac\_layervec\_sitevec}};

    \node[draw,
        minimum width=2cm,
        minimum height=1cm,
        align=center] 
        at (3.6,-2.2) (block3c) 
        {$\Gamma_\textnormal{v}$ in Majorana modes \\
        \texttt{system.gamma\_maj\_sys\_vec}};

    \node[draw,
        minimum width=2cm,
        minimum height=1cm,
        align=center] 
        at (3.6,-3.2) (block4c) 
        {$A, B, D$ submatrices \\
        \texttt{system.mat\_a\_vec}\\
        \texttt{system.mat\_b\_vec} \\
        \texttt{system.mat\_d\_vec} };

    \node[draw,
        minimum width=2cm,
        minimum height=1cm,
        align=center] at (2,-4.6) (block4b) {$\Gamma$ \\ 
        \texttt{system.ferm\_covmat\_vec}};

    \draw[-latex] (block0c) -l (block1c)
        node[pos=0.5,fill=white,inner sep=0]{\bubble{\ref{eq:t-mat-example}}};

    \draw[-latex] ($(block1a.south)!0.82!(block1a.south east)$) -l (block3a){};

    \draw[-latex]
        (block1b) -- (block3a);

    \draw[-latex] (block1c) -l (block2c)
        node[midway, align=center] {\bubble{\ref{eq:gamma_dirac}}};
    \draw[-latex] (block2c) -l (block3c)
        node[midway, align=center]{\bubble{\ref{eq:dirac-majorana-modes}} \\ \bubble{\ref{eq:expand_to_system}}};
    \draw[-latex] (block3c) -l (block4c)
        node[pos=0.5,fill=white,inner sep=0]{};

    \draw[-latex] (block3a) -l (block4b)
        node[pos=0.5,fill=white,inner sep=0]{\bubble{\ref{eq:physical-fermions-cov}}};
    \draw[-latex] (block4c) -l (block4b)
        node[pos=0.5,fill=white,inner sep=0]{\bubble{\ref{eq:physical-fermions-cov}}};
 
\end{tikzpicture}
  \caption{A visualization of the data flow from the parameters describing the state to observable calculation. 
  The circled numbers reference the equation describing the given transition; the monospace font gives the name of the variable relative to a system object.
  In the case of $\Gamma'_\text{in}(\vec{x})$, the indicated variable is actually a dictionary containing $\Gamma'_\text{in}$ for links in each direction.}
  \label{fig:data_flow}
\end{figure}
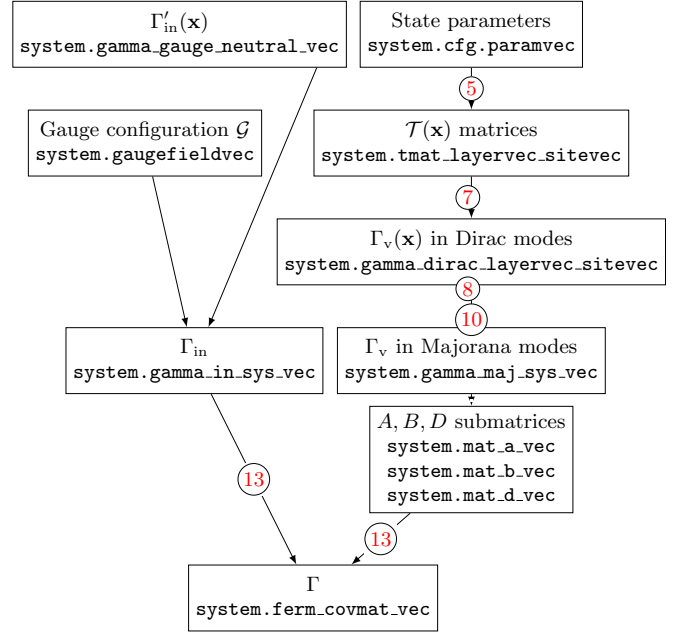

\Cref{fig:data_flow} shows, in graphical form, how the data is handled in the algorithm.
The important transformations are labeled by equation numbers referencing the main 
text.

\section{Monte Carlo Sampling} \label{sec:mc}
The main goal of physical simulations is to predict the expectation values of certain observables.
Given the form of \cref{eq:ansatz-integral-form}, the size of the superposition grows either exponentially with the lattice size for discrete groups or is infinite for continuous groups.
Thus, we cannot expect to evaluate the state exactly for arbitrary parameter values.
Although it is possible to develop similar superpositions in orders~\cite{bender_real-time_2020}, the more general framework is variational Monte Carlo~\cite{sorella_wave_2005}.
Since the concrete structure of the terms in the sum depends on the gauge group, we use variational Monte Carlo.

In contrast with action-based Monte Carlo simulations~\cite{creutz_monte_1979,creutz_monte_1983}, which approximate the partition sum of the system, variational Monte Carlo (VMC) is based on a family of ansatz states based on a set of parameters.
The goal of the procedure is to evaluate the energy and its gradients with respect to the parameters.
These values are subsequently fed into a minimizer to find the parameters minimizing the energy.
Since variational Monte Carlo uses positive semi-definite quantities as Monte Carlo weights, it cannot suffer from the sign problem~\cite{troyer_computational_2005}.
Commonly, the norm of the variational state is used to encode the Monte Carlo weight, i.e. $p(\mathcal{G})=\norm{\psi_\alpha(\mathcal{G}}^2$.
However, the success of a VMC procedure heavily depends on the structure of the Ansatz state.
If it does not have significant overlap or does not include the ground state, the difference in the structure of the ground state and the optimal ansatz state can be significant.
More importantly, a good approximation in energy does not necessarily imply a good convergence of other observables~\cite{sorella_wave_2005}.
Further tests, such as a check of the variance of the energy or evaluation of independent observables, are necessary.

The choice of the Monte Carlo procedure has significant impact on the efficiency of the simulation.
Our aim is to sample the gauge field states $\ket{\mathcal{G}}$ ergodically with an importance sampling procedure.
The simplest and most straightforward update is the update of a single gauge field, corresponding to a single-spin update in a spin-based simulation.
After randomly choosing a gauge field, we update it and then check the acceptance of the Monte Carlo step:
\begin{align}
    p(\mathcal{G} \rightarrow \mathcal{G'})=\min\left(1,p(\mathcal{G})/p(\mathcal{G'})\right).
\end{align}
The gauge field configuration $\mathcal{G'}$ is accepted for sure if it improves the weight.
Otherwise, acceptance is suppressed.
In the code, the logarithm of the squared norm which encodes the weight $p(\mathcal{G})$ is stored instead of the actual value.
Since the squared norm depends on matrices that scale with the system, the norm can vary over multiple orders of magnitude.
Storing the weight as a logarithm prevents loss of precision in floating point operations spanning many orders of magnitude.
The strategy used to choose the new value of the selected gauge field will influence the convergence behavior.
If we choose the value completely randomly, the acceptance probability can be rather low, especially when the group is continuous or contains many discrete elements.
Using the old value of the gauge field to choose a value close by, e.g. a phase close to the previous one in a $U(1)$ setting, will increase the acceptance probability, but it will also increase the correlation between successive steps.

Updating only a single gauge field at a time is not optimal.
As the system grows, only a vanishing fraction of the system is updated in a Monte Carlo step, leading to large auto-correlation times (for details, see Ref.~\cite{gomelski_algorithmic_2025}).
This problem can be mitigated by updating an extensive number of update steps in each step.
One option to achieve an extensive update is the introduction of intermediate update steps which are not followed by a measurement. 
Instead of choosing one single degree of freedom to update in a given step, the algorithm iterates over all d.o.f. and proposes a new configuration.
By possibly taking multiple passes through the system between measurement steps, the autocorrelation time can be reduced.
In Ref.~\cite{gomelski_algorithmic_2025}, we explored different update schemes.

In contrast to action-based MC schemes, the computation of the weight involves the whole system in the GGPEPS construction (see \Cref{eq:norm}).
Thus, it becomes important to optimize the calculation of the norm across the different update steps.
It is performed in every step, whether measurements are taken or not.
The two main operations in \Cref{eq:norm}, are the determinant computation and the inverse of its argument.
Since the gauge field configuration changes only on a single link in both, the heatbath and the single-spin update, we can propagate information from the previous step to the current one.
This propagation happens via the matrix determinant lemma~\cite{harville_matrix_1997} and the Woodbury inversion formula.
This is more efficient than straightforwardly multiplying the system-sized matrices 
at every step; $D$ does not depend on $\mathcal{G}$ and so its determinant need only 
be calculate once during the MC procedure.

The update procedure happens in two distinct steps:
first, the inverse is updated using the Woodbury inversion formula
\begin{equation} 
    \label{eq:inv-update}
    \begin{aligned}
    &(M+UCV)^{-1} \\
        &= M^{-1} - M^{-1}U(C^{-1}+VM^{-1}U)^{-1} VM^{-1} \\
        &= M^{-1} - M^{-1}U (\id + CVM^{-1}U)^{-1} CVM^{-1},
\end{aligned} \end{equation}
where $M,U,C,V$ are invertible matrices of the appropriate size.
The two formulations of the lemma are mathematically equivalent, however, they are performing differently numerically.
In the third line, the inverse of the update matrix $C$ is avoided.
Thus, the update can also be performed for singular updates as they can occur in non-Abelian gauge groups like $D_6$.
In the code, the matrix inversion lemma is used to compute $(D^{-1} - \Gamma_{\text{in}}(\mathcal{G}))^{-1}$, a part of the norm computation and $(D - \Gamma_{\text{in}}(\mathcal{G}))^{-1}$.
The matrix $C$ is the update to the current $\Gamma_{\text{in}}(\mathcal{G})$ which is positioned by $U$ and $V$.
In a second step, the tracked determinant is updated using the new inverse of $A$
\begin{equation} \begin{aligned}
    \label{eq:det-update}
    \det(M+UCV) 
        &= \det(M) \det(C^{-1} + VM^{-1}U) \det(C) \\
        &= \det(M) \det(\id + V M^{-1} U C).
\end{aligned} \end{equation}
Together, these two update procedures avoid computing the repeated inverse and determinant of a system-sized matrix.

In the long-term, it would be interesting to find genuinely global updates for the code.
In spin-systems, Wolff- \cite{wolff_collective_1989} and Swendsen-Wang style~\cite{swendsen_nonuniversal_1987} or more general loop updates are possibilities to change an extensive fraction of the system (see~\cite{sandvik_computational_2010} and references therein).
However, designing similar updates for continuous gauge groups is rather involved.
Additional considerations involve including Langevin ~\cite{aarts_can_2009} or hybrid Monte Carlo schemes~\cite{duane_hybrid_1987, sexton_hamiltonian_1992, hasenbusch_speeding_2001} to find better updates for the gauge fields.

\section{Code}
After the more technical introduction into the structure of the states, we now consider the practical use of the \ggpeps package.
Many of the subtleties discussed above are abstracted away and need not be considered when running typical simulations.
We first describe the setup of an ansatz state, then its use in evaluating observables for fixed variational parameters, and finally describe a ground state search.

\subsection{Code: Structure and Use}
\label{sec:code}

To install the package, clone the repository to an appropriate directory, and then run 
\begin{Verbatim}[commandchars=\\\{\},fontsize=\small,breaklines]
pip\PYG{+w}{ }install\PYG{+w}{ }.
\end{Verbatim}

or, if you intend to edit the code, mark the code as editable by installing with
\begin{Verbatim}[commandchars=\\\{\},fontsize=\small,breaklines]
pip\PYG{+w}{ }install\PYG{+w}{ }\PYGZhy{}e\PYG{+w}{ }.
\end{Verbatim}

More detailed instructions can be found in the project's README~\cite{Kelman-GGPEPSGithubSource-2026}.
The code below was tested on commit \texttt{83c30da}. 

\begin{figure}
  \centering
  \begin{tikzpicture}[scale=1.5]


    \node[draw,
        minimum width=2cm,
        minimum height=1cm,
        align=center] at (-1,-1) (block2a) 
        {Minimizer \\ \texttt{ggpeps.Minimizer}};

    \node[draw,
        minimum width=2cm,
        minimum height=1cm,
        align=center] at (1,-2) (block2b) 
        {EvaluatorManager \\
        \texttt{ggpeps.EvaluatorManager}};

    \node[draw,
        minimum width=2cm,
        minimum height=1cm,
        align=center] at (1,-3) (block3) 
        {Evaluator(s) \\ \texttt{ggpeps.Evaluator}};

    \node[draw,
        minimum width=2cm,
        minimum height=1cm,
        align=center] at (1,-4) (block4) 
        {system \\ \texttt{ggpeps.system.System2DBase}};

    \node[draw,
        minimum width=2cm,
        minimum height=1cm,
        align=center] at (1,-5) (block5) 
        {system config \\ \texttt{ggpeps.system.Config2DBase}};

    \node[draw,
        minimum width=2cm,
        minimum height=1cm,
        align=center] at (-0.3,-6) (block6a) 
        {gauge group \\ \texttt{ggpeps.GaugeGroup}};

    \node[draw,
        minimum width=2cm,
        minimum height=1cm,
        align=center] at (2.3,-6) (block6b) 
        {lattice \\ \texttt{ggpeps.Lattice2D}};



    \draw[-latex] (block2a) -l (block2b) {};

    \draw[-latex] (block2b) -l (block3) {};

    \draw[-latex] (block3) -l (block4) {};

    \draw[-latex] (block4) -l (block5) {};

    \draw[-latex] (block5) -l (block6a) {};
    \draw[-latex] (block5) -l (block6b) {};

\end{tikzpicture}
  \caption{Ownership structure of objects in the code.
  The monospace font indicates the type of the object (in some cases, this is an abstract base class).}
  \label{fig:object_structure}
\end{figure}
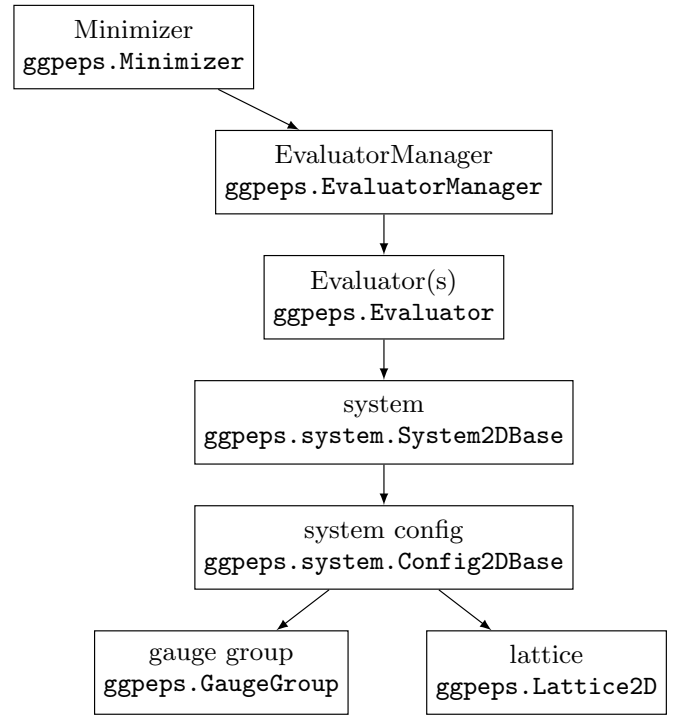

Now we turn to describing use of the code. 
The ownership structure of the objects discussed in this section is illustrated in figure~\ref{fig:object_structure}.

The \texttt{system} object resides at a low level of the code. 
This object implements the quantum state $\ket{\psi_\alpha(\mathcal{G})}$. 
That is, it contains an attribute which holds the current gauge configuration, as well as all the quantities and operations needed to calculate any observable on the state $\ket{\psi_\alpha(\mathcal{G})}$.
At present, the system is defined as a child class (unique to a given gauge group) of a parent system base class, which implements the functionality shared between all gauge groups (including much of the logic of figure~\ref{fig:data_flow}).

Creating a system requires a \texttt{system\_config} object which defines the ansatz --- the gauge group, number of virtual modes, Hamiltonian couplings, and other system settings.
Both the \texttt{system} and \texttt{system\_config} are defined using an abstract base class, with each system (config) inheriting from the appropriate base.

Thus, the first step in running a simulation is specifying the system configuration --- the Hamiltonian couplings, size of the lattice, as well as the specification of the ansatz, such as the number of copies.
First, we import the required modules:
\begin{Verbatim}[commandchars=\\\{\},fontsize=\small,breaklines]
\PYG{k+kn}{import}\PYG{+w}{ }\PYG{n+nn}{numpy}\PYG{+w}{ }\PYG{k}{as}\PYG{+w}{ }\PYG{n+nn}{np}
\PYG{k+kn}{from}\PYG{+w}{ }\PYG{n+nn}{ggpeps}\PYG{+w}{ }\PYG{k+kn}{import} \PYG{n}{lattice} \PYG{k}{as} \PYG{n}{lat}
\PYG{k+kn}{from}\PYG{+w}{ }\PYG{n+nn}{ggpeps}\PYG{n+nn}{.}\PYG{n+nn}{minimizer}\PYG{+w}{ }\PYG{k+kn}{import} \PYG{n}{MinimizerConfig}\PYG{p}{,} \PYG{n}{Minimizer}
\PYG{k+kn}{from}\PYG{+w}{ }\PYG{n+nn}{ggpeps}\PYG{n+nn}{.}\PYG{n+nn}{evaluator\PYGZus{}manager}\PYG{+w}{ }\PYG{k+kn}{import} \PYG{n}{EvaluatorManager}
\PYG{k+kn}{from}\PYG{+w}{ }\PYG{n+nn}{ggpeps}\PYG{n+nn}{.}\PYG{n+nn}{mc}\PYG{+w}{ }\PYG{k+kn}{import} \PYG{n}{MonteCarloEvaluatorConfig}
\PYG{k+kn}{from}\PYG{+w}{ }\PYG{n+nn}{ggpeps}\PYG{n+nn}{.}\PYG{n+nn}{system}\PYG{+w}{ }\PYG{k+kn}{import} \PYG{n}{Z2System2D\PYGZus{}Config}\PYG{p}{,} \PYG{n}{Z2System2D}
\end{Verbatim}

and then create a lattice:
\begin{Verbatim}[commandchars=\\\{\},fontsize=\small,breaklines]
\PYG{n}{L} \PYG{o}{=} \PYG{l+m+mi}{2} \PYG{c+c1}{\PYGZsh{} size of the lattice}
\PYG{n}{gauge\PYGZus{}fixing} \PYG{o}{=} \PYG{o}{\PYGZhy{}}\PYG{l+m+mi}{1} \PYG{c+c1}{\PYGZsh{} maximal tree}
\PYG{n}{lattice} \PYG{o}{=} \PYG{n}{lat}\PYG{o}{.}\PYG{n}{Lattice2D}\PYG{p}{(}\PYG{n}{L}\PYG{p}{,} \PYG{n}{L}\PYG{p}{,} \PYG{n}{gauge\PYGZus{}fixing}\PYG{p}{)}
\end{Verbatim}

The created lattice will have periodic boundary conditions~\cite{gomelski_algorithmic_2025}.

Next, we create a system configuration object. 
We specify the ansatz parameters as well as the Hamiltonian couplings:
\begin{Verbatim}[commandchars=\\\{\},fontsize=\small,breaklines]
\PYG{n}{system\PYGZus{}cfg} \PYG{o}{=} \PYG{n}{Z2System2D\PYGZus{}Config}\PYG{p}{(}
    \PYG{n}{lattice}\PYG{p}{,} 
    \PYG{n}{g\PYGZus{}el}\PYG{o}{=}\PYG{l+m+mf}{1.0}\PYG{p}{,}
    \PYG{n}{g\PYGZus{}mag}\PYG{o}{=}\PYG{l+m+mf}{1.0}\PYG{p}{,}
    \PYG{n}{g\PYGZus{}int}\PYG{o}{=}\PYG{l+m+mf}{1.0}\PYG{p}{,}
    \PYG{n}{g\PYGZus{}mass}\PYG{o}{=}\PYG{l+m+mf}{1.0}\PYG{p}{,}
    \PYG{n}{g\PYGZus{}chem}\PYG{o}{=}\PYG{p}{[}\PYG{l+m+mf}{0.0}\PYG{p}{]}\PYG{p}{,}
    \PYG{n}{ncopy}\PYG{o}{=}\PYG{l+m+mi}{2}\PYG{p}{,}
    \PYG{n}{num\PYGZus{}pg\PYGZus{}layer}\PYG{o}{=}\PYG{l+m+mi}{1}\PYG{p}{,}
    \PYG{n}{num\PYGZus{}fermionic\PYGZus{}layer}\PYG{o}{=}\PYG{l+m+mi}{1}\PYG{p}{,}
    \PYG{n}{mod\PYGZus{}link\PYGZus{}inds}\PYG{o}{=}\PYG{p}{(}\PYG{l+m+mi}{0}\PYG{p}{,} \PYG{l+m+mi}{1}\PYG{p}{)}\PYG{p}{,}
    \PYG{n}{unitcell\PYGZus{}size}\PYG{o}{=}\PYG{l+m+mi}{1}\PYG{p}{,}
    \PYG{n}{enforce\PYGZus{}u1\PYGZus{}symmetry}\PYG{o}{=}\PYG{k+kc}{True}\PYG{p}{,}
\PYG{p}{)}
\end{Verbatim}

and then set the variational parameters (for example, to random parameters),
\begin{Verbatim}[commandchars=\\\{\},fontsize=\small,breaklines]
\PYG{n}{shape} \PYG{o}{=} \PYG{n}{system\PYGZus{}cfg}\PYG{o}{.}\PYG{n}{param\PYGZus{}shape}\PYG{p}{(}\PYG{p}{)}
\PYG{n}{params} \PYG{o}{=} \PYG{n}{np}\PYG{o}{.}\PYG{n}{random}\PYG{o}{.}\PYG{n}{RandomState}\PYG{p}{(}\PYG{l+m+mi}{42}\PYG{p}{)}\PYG{o}{.}\PYG{n}{rand}\PYG{p}{(}\PYG{o}{*}\PYG{n}{shape}\PYG{p}{)}
\PYG{n}{system\PYGZus{}cfg}\PYG{o}{.}\PYG{n}{paramvec} \PYG{o}{=} \PYG{n}{params}
\end{Verbatim}

Finally, we create a system object
\begin{Verbatim}[commandchars=\\\{\},fontsize=\small,breaklines]
\PYG{n}{system\PYGZus{}type} \PYG{o}{=} \PYG{n}{Z2System2D}
\PYG{n}{system} \PYG{o}{=} \PYG{n}{system\PYGZus{}type}\PYG{p}{(}\PYG{n}{system\PYGZus{}cfg}\PYG{p}{)}
\end{Verbatim}

The \texttt{system} object is set up to capture the state $\ket{\psi_\alpha(\mathcal{G})}$, with all details specific to a particular gauge group or ansatz setup delegated to the \texttt{system\_config}.
If we wish to represent the state $\ket{\psi_\alpha(\mathcal{G})}$ for some particular $\mathcal{G}$, we can update the system to match (by default, it starts with the all-identity configuration).
We can access the allowed gauge group values
\begin{Verbatim}[commandchars=\\\{\},fontsize=\small,breaklines]
\PYG{n}{allowed\PYGZus{}gauges} \PYG{o}{=} \PYG{n}{system}\PYG{o}{.}\PYG{n}{cfg}\PYG{o}{.}\PYG{n}{gaugemgr}\PYG{o}{.}\PYG{n}{get\PYGZus{}possible\PYGZus{}gauge\PYGZus{}values}\PYG{p}{(}\PYG{p}{)}
\PYG{n}{nlinks} \PYG{o}{=} \PYG{n}{system}\PYG{o}{.}\PYG{n}{cfg}\PYG{o}{.}\PYG{n}{lattice}\PYG{o}{.}\PYG{n}{nlinks}
\PYG{n}{gauge\PYGZus{}configuration} \PYG{o}{=} \PYG{p}{[}\PYG{n}{allowed\PYGZus{}gauges}\PYG{p}{[}\PYG{l+m+mi}{0}\PYG{p}{]}\PYG{p}{]}\PYG{o}{*}\PYG{n}{nlinks}
\end{Verbatim}

of course, any other choice of gauge configuration is equally acceptable.
The goal of the Monte Carlo procedure is to sample the different configurations according to their weight.
Thus, all gauge configurations must be representable.

To evaluate an observable on this gauge configuration, update the gauge configuration of the system, and then perform the desired computation. 
For example
\begin{Verbatim}[commandchars=\\\{\},fontsize=\small,breaklines]
\PYG{n}{system}\PYG{o}{.}\PYG{n}{update\PYGZus{}gauge\PYGZus{}full\PYGZus{}system}\PYG{p}{(}\PYG{n}{gauge\PYGZus{}configuration}\PYG{p}{)}
\PYG{n}{mag\PYGZus{}energy} \PYG{o}{=} \PYG{n}{system}\PYG{o}{.}\PYG{n}{mag\PYGZus{}energy}
\end{Verbatim}

If we wish to evaluate observables, we must perform calculations on the state $\ket{\psi_\alpha}$, and not just on a single term for a fixed gauge configuration $\ket{\psi_\alpha(G)}$.
To do so, we create an \texttt{EvaluatorManager}. 
The \texttt{EvaluatorManager} is a wrapper around \texttt{Evaluator} objects, each of which handle states (i.e. a \texttt{system} object) for multiple gauge configurations. 
This structure is useful, as it allows splitting the work between several \texttt{Evaluator}s, and ultimately between several computational cores. 
The prime example of this is running multiple \texttt{MonteCarloEvaluator}s in parallel.

Thus an \texttt{EvaluatorManager} estimates the expectation values with respect to the full state $\ket{\psi_\alpha} = \int \mathcal{DG} \ket{\psi_\alpha(\mathcal{G})}$ for all observables of interest.
The code currently has two evaluator types: an \texttt{ExactEvaluator} which computes the expectation values by explicitly summing over all gauge configurations $\mathcal{G}$, and a \texttt{MonteCarloEvaluator} which estimates the same (and produces error estimates) based on a Markov Chain Monte Carlo procedure.
The exact evaluator can only be used for discrete gauge groups, and the calculation quickly becomes infeasible as the number of gauge configurations grows (due either to the size of the gauge group or the size of the lattice and resulting growth of the Hilbert space).

To run an evaluation, we start by specifying a configuration --- in the case of Monte Carlo, this includes settings such as the number of steps to take in an evaluation.
With that configuration, we create an \texttt{EvaluatorManager} by specifying the system type, the ansatz configuration, and the evaluator configuration:
\begin{Verbatim}[commandchars=\\\{\},fontsize=\small,breaklines]
\PYG{n}{mc\PYGZus{}config} \PYG{o}{=} \PYG{n}{MonteCarloEvaluatorConfig}\PYG{p}{(}
        \PYG{n}{warmup\PYGZus{}steps}\PYG{o}{=}\PYG{l+m+mi}{1000}\PYG{p}{,}
        \PYG{n}{meas\PYGZus{}steps}\PYG{o}{=}\PYG{l+m+mi}{1000}\PYG{p}{,}
        \PYG{n}{compute\PYGZus{}grads}\PYG{o}{=}\PYG{k+kc}{False}\PYG{p}{,}
        \PYG{n}{update\PYGZus{}size\PYGZus{}per\PYGZus{}step}\PYG{o}{=}\PYG{l+m+mi}{1}
    \PYG{p}{)}
\PYG{n}{eval\PYGZus{}manager} \PYG{o}{=} \PYG{n}{EvaluatorManager}\PYG{p}{(}\PYG{n}{system\PYGZus{}type}\PYG{p}{,} \PYG{n}{system\PYGZus{}cfg}\PYG{p}{,} \PYG{n}{mc\PYGZus{}config}\PYG{p}{)}
\end{Verbatim}

As mentioned, the code supports running several Monte Carlo Evaluators in parallel, which is managed by the \texttt{EvaluatorManager} when provided with the addition \texttt{nrunner} argument. 
We pass the \texttt{system} class itself, rather than an instance of it, to allow the \texttt{EvaluatorManager} to instantiate a \texttt{system} for each runner.
This uses several \texttt{Evaluator} objects in parallel and merges the results at the end. 
This is useful when many computational cores are available, but at present, the efficiency gains of using many runners depend on implementation details and the system in question, which differ in which operations are more/less expensive. 
Before running a lengthy simulations with multiple runners, we recommend some benchmarking to determine the optimal way to split up the number of steps among some number of runners.

When one wishes to calculate observables on a state defined by known parameters, the \texttt{EvaluatorManager} object is the root object for a simulation.
In such a case, the code
\begin{Verbatim}[commandchars=\\\{\},fontsize=\small,breaklines]
\PYG{n}{eval\PYGZus{}manager}\PYG{o}{.}\PYG{n}{simulate}\PYG{p}{(}\PYG{p}{)}
\PYG{n}{mc\PYGZus{}result} \PYG{o}{=} \PYG{n}{eval\PYGZus{}manager}\PYG{o}{.}\PYG{n}{get\PYGZus{}evaluator}\PYG{p}{(}\PYG{p}{)}
\PYG{n}{mc\PYGZus{}result}\PYG{o}{.}\PYG{n}{print\PYGZus{}stats}\PYG{p}{(}\PYG{p}{)}
\PYG{n}{mc\PYGZus{}result}\PYG{o}{.}\PYG{n}{save}\PYG{p}{(}\PYG{l+s+s2}{\PYGZdq{}}\PYG{l+s+s2}{\PYGZlt{}output\PYGZus{}dir\PYGZgt{}}\PYG{l+s+s2}{\PYGZdq{}}\PYG{p}{)}
\end{Verbatim}

calculates all of the defined observables (in \texttt{simulate()}), prints a summary of the results, and then saves the result to disk.
Note that since gradients are computationally expensive, in order to compute them, one must set 
\begin{Verbatim}[commandchars=\\\{\},fontsize=\small,breaklines]
\PYG{n}{mc\PYGZus{}config}\PYG{o}{.}\PYG{n}{compute\PYGZus{}grads} \PYG{o}{=} \PYG{k+kc}{True}
\end{Verbatim}

before running the simulation.

Often the code is run in minimization mode, in which repeated evaluations are done in order to optimize the parameters, e.g. in order to find the ground state.
This is handled by a \texttt{Minimizer} object, which owns the \texttt{EvaluatorManager}.
This supports many of \scipy's minimization methods, as well as several custom minimizers.
\begin{Verbatim}[commandchars=\\\{\},fontsize=\small,breaklines]
\PYG{n}{min\PYGZus{}cfg} \PYG{o}{=} \PYG{n}{MinimizerConfig}\PYG{p}{(}\PYG{n}{method}\PYG{o}{=}\PYG{l+s+s2}{\PYGZdq{}}\PYG{l+s+s2}{BFGS}\PYG{l+s+s2}{\PYGZdq{}}\PYG{p}{)}
\PYG{n}{minimizer} \PYG{o}{=} \PYG{n}{Minimizer}\PYG{p}{(}\PYG{n}{min\PYGZus{}cfg}\PYG{p}{,} \PYG{n}{eval\PYGZus{}manager}\PYG{p}{)}
\PYG{n}{result} \PYG{o}{=} \PYG{n}{minimizer}\PYG{o}{.}\PYG{n}{minimize}\PYG{p}{(}\PYG{p}{)}
\end{Verbatim}

Depending on the minimization algorithm used, gradients will automatically be computed if (and only if) necessary.

Results can be saved to disk by calling the \texttt{save()} functions of the minimizer and evaluator. 
Additionally, the code makes use of a logger at various levels. 
By default, the logging level is \texttt{WARNING} and prints to \texttt{stderr}; the logger can be set up by calling \texttt{ggpeps.utils.setup\_logger()}, or further customization can be done using the \texttt{logging} module in python.

As mentioned throughout this section, determining the optimal settings is not a trivial task. 
This is true of the ansatz settings --- for instance, how many copies of virtual modes to include, but equally true for the evaluation of observables --- e.g. the number of Monte Carlo steps, and the minimization. 

For some of the objects mentioned here, there are a variety of settings which we have not mentioned here, which may be useful but distract from the main structure and goal of this paper. We refer the interested reader to the README and to the code itself~\cite{Kelman-GGPEPSGithubSource-2026}.

\begin{figure}[t]
    \centering
    \includegraphics[width=0.95\linewidth]{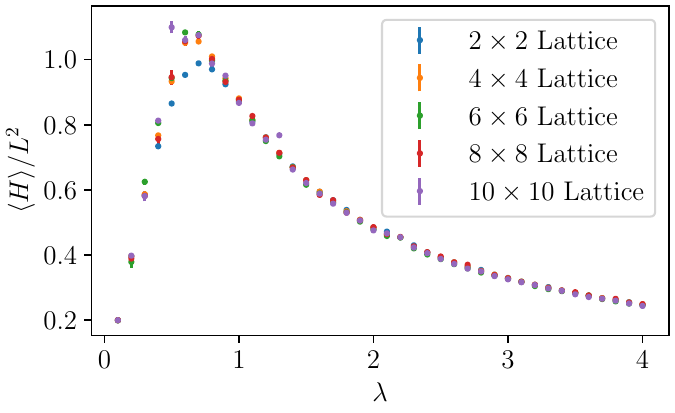}
    \caption{Pure gauge results for a $\mathbb{Z}_2$ theory of varying lattice sizes. Results show the ground state energy found with a GGPEPS ground state search with varying coupling parameter $\lambda$. See~\Cref{sec:z2-pg-system} for details.}
    \label{fig:z2-pg}
\end{figure}

For previous specific examples of results produced by our code, we refer the reader to figure 6 of~\cite{emonts_finding_2023} showing the ground state energy for a $\mathbb{Z}_2$ pure-gauge theory. 
Ground state energy results for a $\mathbb{Z}_2$ theory including fermionic matter can be seen in figures 4 and 6 of~\cite{kelman_projected_2026}. 
Figure 7 there shows Wilson loop expectation values on the ground state result, and possibly indicates a phase transition.
As a demonstration of recent development of our code, we also include here, in figure~\ref{fig:z2-pg}, results for the pure gauge $\mathbb{Z}_2$ theory described in~\cite{emonts_finding_2023} for systems up to a lattice size of $10 \times 10$, something which was previously unachievable with this method. We summarize the system in~\Cref{sec:z2-pg-system}.

\subsection{Implementation Notes}
\label{sec:implementation_details}

In this section we note some important implementation details. 
Most of these are irrelevant to a user of the code; however, the note on JAX/numpy is necessary to run the code on GPUs.

Many system attributes are evaluated lazily and on demand, with caching to avoid expensive recomputation unless necessary. 
This also avoids the need to consider the order in which quantities are computed --- using any quantity will automatically calculate any quantity on which it depends, all of which will then be cached. 
The general pattern looks like:
\begin{Verbatim}[commandchars=\\\{\},fontsize=\small,breaklines]
\PYG{k}{class}\PYG{+w}{ }\PYG{n+nc}{System}\PYG{p}{:}
    \PYG{n+nd}{@property}
    \PYG{k}{def}\PYG{+w}{ }\PYG{n+nf}{some\PYGZus{}quantity}\PYG{p}{(}\PYG{n+nb+bp}{self}\PYG{p}{)}\PYG{p}{:}
        \PYG{k}{if} \PYG{n+nb+bp}{self}\PYG{o}{.}\PYG{n}{\PYGZus{}some\PYGZus{}quantity} \PYG{o+ow}{is} \PYG{k+kc}{None}\PYG{p}{:}
            \PYG{n+nb+bp}{self}\PYG{o}{.}\PYG{n}{\PYGZus{}some\PYGZus{}quantity} \PYG{o}{=} \PYG{n+nb+bp}{self}\PYG{o}{.}\PYG{n}{compute\PYGZus{}some\PYGZus{}quantity}\PYG{p}{(}\PYG{p}{)}
        \PYG{k}{return} \PYG{n+nb+bp}{self}\PYG{o}{.}\PYG{n}{\PYGZus{}some\PYGZus{}quantity}
\end{Verbatim}

By using python properties, the true attributes starting with an underscore (\texttt{\_some\_quantitity}) are only accessible via the functions generated by the property decorator.
This pattern ensures that the expensive function \texttt{compute\_some\_quantity()} is only called once.

The \ggpeps code is set up to run with two different backends: \numpy and \jax, taking advantage of their similar syntax. 
The choice between \numpy and \jax is controlled by an environment variable, \texttt{GGPEPS\_BACKEND},  loaded at startup.
Each of \numpy and \jax have advantages and disadvantages; most relevant are
\begin{itemize}
    \item Support for just-in-time compilation and GPUs in \jax.
    \item Runtime: on CPUs, our benchmarking shows that numpy is often faster, but \jax is faster for larger systems when using just-in-time compilation.
    \item Code flexibility: certain \jax requirements, especially when using just-in-time compilation, place restrictions on allowed operations (e.g. allowed control flow, such as \texttt{if-else} statements). 
    More information can be found in the \jax documentation~\cite{bradbury_jax_2018}. 
    This has given rise to a class-based structure in our code which borrows heavily from a functional paradigm; many of the heavy computations are done in \texttt{jit}-compiled functions, making use of the \texttt{@staticmethod} decorator.
\end{itemize}

In certain cases, such as array assignments or Pfaffian calculations, different syntax is needed for \numpy and \jax. 
This is handled via a wrapper \texttt{Backend} class (chosen upon import of the \texttt{ggpeps} package), which implements the needed functionality through a unified interface, and which is chosen at startup when the choice between \numpy and \jax is made.
In either case, results are saved as \numpy data types; this setting only affects data within the \texttt{system} object.

The code contains many multi-dimensional arrays, and it is crucial to ensure consistency between operations and track the structure (e.g. mode order) of each array. 
At the highest level, arrays often include a dimension for the layer and site. 
The innermost dimensions are generally those which correspond to the degrees of freedom; i.e. they are (covariance) matrices whose dimension is determined by the number of degrees of freedom in the system. 
These include physical and virtual modes (but not the gauge degrees of freedom --- the matrices depend on the gauge configuration, but their dimensionality does not).
It is sometimes convenient to work with site-based mode ordering, and other times more convenient to work link-based mode ordering --- swapping between the two is handled by an automatic permutation builder.
This is, of course, a crucial detail --- any matrix operation must ensure that the mode ordering is compatible.
For this purpose, the code includes utilities that convert from (for example) a site-based mode order to a link-based one, and that can generate permutation matrices between arbitrary orders.

\subsection{Tests}
The code comes with an extensive testing suite.
The main suite uses python's \texttt{unittest}, and can be run with \texttt{python -m unittest}. 
There is also a \texttt{nox} suite which is used to test builds, linting, and type hints, as well as to run the testing suite with both the \numpy and \jax backends. The full \texttt{nox} suite can be run with the simple command \texttt{nox}, and invididual sessions (e.g. \texttt{mypy} typing) can be run with \texttt{nox -s typing}.

\section{Conclusion and Outlook}

GGPEPS provides an ansatz tailored for lattice gauge theories: satisfying 
desired symmetries, entanglement properties, and avoiding the sign problem 
while remaining computationally tractable. In previous work we have 
extensively developed the construction and properties of the ansatz, and 
elsewhere demonstrated numerical results for a variety of system. Here we 
presented the code used in computing those results, with a guide to its structure and use. As the code continues to be developed, this will enable 
greater transparency and auditing, as well as use of the GGPEPS method in 
other contexts.

There are a variety of ways one could extend the code; this subsection provides guidance on implementing some of the possible extensions.
One direction for extensions is adding additional observables. 
This involves implementing the observable in terms of the quantities defined in figure~\ref{fig:data_flow}, or in terms of quantities derivable from them. 
This can be done in the system base class, if the implementation can be shared between different gauge groups, or the implementation can be unique to each gauge group.
After the observable is implemented, the evaluators must also be updated to measure the observable.

Another major direction for improvement is adding support for additional gauge groups; currently support for the Dihedral groups $D_n$ (i.e. a simple family of non-Abelian groups) is in progress. 
Adding support for a new gauge group involves three steps: defining the gauge group, defining the ansatz, and defining the system. 
To define the gauge group, one must create a child class of the \texttt{GaugeGroup} parent class.
To define the ansatz, one must subclass the \texttt{config\_base} class. Some analytical work is required to do this, such as choosing a parameterization of the $\mathcal{T}$ matrices that obeys the desired symmetries.
The last of these requires subclassing the \texttt{system} base class, and implementing the abstract methods, such as the function \texttt{system.\_update\_gauge\_ind()} which performs the gauging operation.

Porting the code to GPUs and vectorizing crucial methods has given more than an order of magnitude speedup;
applying GGPEPS methods to larger or more complex systems will likely require further algorithmic and computational optimizations, as well as expertise in the relevant physics.
In particular, greater computational efficiency will enable the extending of the ansatz to 3 spatial dimensions~\cite{emonts_fermionic_2023}. 
Apart from the theoretical work including the correct account for spin, the generalization of the code to three spatial dimensions is possible in the current framework.

\acknowledgments
We gratefully acknowledge the many people who have worked on the theoretical development of the GGPEPS ansatz, as well as those who have used or contributed, directly or indirectly, to the \texttt{ggpeps} package.
This includes (in alaphabetical order) Mari Carmen Bañuls, Michele Burrello, J. Ignacio Cirac, Uri Friedman, Gertian Roose, Thorsten B. Wahl, Harel Wullman.

In developing this code, we have run simulations on several computing clusters, including the Fritz Haber Center for Molecular Dynamics at Hebrew University, the Academic Leiden Interdisciplinary Cluster Environment (ALICE) provided by Leiden University, and the EU Large Unified Modern Infrastructure (LUMI) supercomputer. 
These have provided the opportunity to test, debug, and optimize on a variety of hardware platforms (including GPUs).

This research was funded  by the European Union (ERC, OverSign, 101122583). 
Views and opinions expressed are those of the authors only and do not necessarily reflect those of the European Union or the European Research Council.

P.E. acknowledges the support received from the Dutch National Growth Fund (NGF) as part of the Quantum Delta NL program in the NWO-Quantum Technology program (Grant No.~NGF.1623.23.006).
P.E. also acknowledges funding from the Carl-Zeiss-Stiftung (CZS Center QPhoton). 

\appendix

\section{Z2 Pure Gauge System}
\label{sec:z2-pg-system}

A pure gauge $\mathbb{Z}_2$ system is instantiated by a lattice as illustrated in~\Cref{fig:lattice}, with periodic boundary conditions.
Each lattice edge (or link) is associated with a spin-1/2 Hilbert space.
No matter is included on the sites.
The dynamics is given by the Hamiltonian~\cite{horn_hamiltonian_1979}
\begin{equation} 
    \begin{aligned}
        H &= \lambda\sum_\ell \left[1-\sigma^z_\ell\right] + \frac{1}{\lambda}\underset{p}{\sum}\left[1-\sigma^x_{p_1} \sigma^x_{p_2} \sigma^x_{p_3} \sigma^x_{p_4}\right].
    \end{aligned} 
    \label{eq:ham}
\end{equation}

The Hamiltonian is gauge invariant; that is, it is invariant under \emph{local} unitary transformations of the form
\begin{equation}
V\left(\xarg\right) =   \sigma^z\left(\xarg,1\right)\sigma^z\left(\xarg,2\right)\sigma^z\left(\xarg-\vu{e}_1,1\right)\sigma^z\left(\xarg-\vu{e}_2,2\right).
\end{equation}
Figure~\Cref{fig:z2-pg} shows ground states found using GGPEPS for this system at varying lattice sizes. For more details on this particular system, see~\cite{emonts_finding_2023}. 

\section{Electric Energy Implementation}
\label{sec:electric_implementation}
The electric energy is arguably the most complicated observable to compute.
This complexity arises because, unlike many other physical observables, it is not diagonal in the group element basis in which we perform the computation of other expectation values.
We present here the computation for a single link $\ell_q$; 
the electric energy of the whole lattice is then obtained by summing over all links, which in the case of a translationally invariant system reduces to multiplying the single-link result by the number of links (when considering Monte Carlo sampling, this increases the error: to achieve a given error it can be beneficial to compute the electric energy over more links for fewer steps, or over fewer links for more steps). 
The code can also evaluate the energy on any chosen subset of links, in which case it simply repeats the single-link procedure described below for each link in the subset.
For a single link $\ell_0$, we evaluate
\begin{equation}
    \left\langle H_E(\ell_0) \right\rangle = \int\mathcal{DG}\ F_E\left(\mathcal{G}\right) p\left(\mathcal{G}\right)
\end{equation}
where $p\left(\mathcal{G}\right)=\frac{\braket{\psi_\alpha\left(\mathcal{G}\right)}{\psi_\alpha\left(\mathcal{G}\right)}}{\int \mathcal{DG}'\braket{\psi\left(\mathcal{G}'\right)}{\psi\left(\mathcal{G}'\right)}}$
is the normalized Monte Carlo weight, the numerator of which can be computed using Eq.~\eqref{eq:norm}.
The function $F_E\left(\mathcal{G}\right)$ is given by 
\begin{equation} \label{eq:FE}
    F_E\left(\mathcal{G}\right) 
    = \int dg' \frac{\braket{\psi_\alpha(\mathcal{G'})}{\psi_\alpha(\mathcal{G})}}{\braket{\psi_\alpha(\mathcal{G})}{\psi_\alpha\left(\mathcal{G}\right)}} \bra{g'} H_E (\ell_0)\ket{g}_{\ell_0},
\end{equation}
where 
\begin{equation} \begin{aligned}
    \mathcal{G} &= (g_1, ..., g_{l_0 - 1}, g, g_{l_0 + 1}, ..., g_{N_\text{links}}) \\
    \mathcal{G}' &= (g_1, ..., g_{l_0 - 1}, g', g_{l_0 + 1}, ..., g_{N_\text{links}}) \\
\end{aligned} \end{equation} 
are the same gauge field configurations but with $g$ or $g'$ on link $\ell = l_0$.
The term $\bra{g'} H_E (\ell_0)\ket{g}_{\ell_0}$ imposes a constraint on the relationship between $g$ and the integration variable $g'$ on the link $\ell_0$.
For finite groups, this constraint typically yields a sum over Kronecker deltas that can be integrated out;
for example, in the $\mathbb{Z}_2$ case, it enforces the condition $g \neq g'$, i.e. that one of them is the identity element, and the other is the only non-identity element of $\mathbb{Z}_2$.

Thus, calculating $F_E(\mathcal{G})$ reduces to computing the overlap
$\braket{\psi_\alpha\left(\mathcal{G} \right)}{\psi_\alpha (\mathcal{G}')}$
for specific group elements $g'$.
This is achieved through the following simplification:
\begin{widetext}
\begin{equation}
    \braket{\psi_\alpha\left(\mathcal{G} \right)}{\psi_\alpha\left(\mathcal{G}'\right)} 
    = \bra{\chi\left(\mathcal{G}\right)} \Big[ \mathcal{U}_h^\dagger(\ell_0) \omega(\ell_0) \ket{\Omega_v(\ell_0)}\bra{\Omega_v(\ell_0)} \omega^\dagger(\ell_0) \Big] \ket{\chi(\mathcal{G})}
\end{equation}
where $h = g^{-1} g'$ and
\begin{equation}
\label{eq:chi-def}
    \ket{\chi\left(\mathcal{G}\right)}=\prod_{l \neq \ell_0} \bra{\Omega_v(l)} \omega^\dagger\left(l\right) \prod_{l'} \mathcal{U}_g(l') \prod_{\vec{x}}A(\vec{x})\ket{\Omega_p}\ket{\Omega_v}.
\end{equation}
\end{widetext}
Note that the projector $\omega$ and the vacuum $\bra{\Omega_v}$ are excluded on the link $\ell_0$.

Expanding the central operator $\mathcal{U}_h^\dagger \omega \ket{\Omega_v}\bra{\Omega_v} \omega^\dagger$ using the definitions of the projector $\omega$ and gauging operator $\mathcal{U}_g$ from Eq.~\eqref{eq:ansatz-components} yields a polynomial in the Majorana fermionic modes $c_j$ on link $\ell_0$,
\begin{widetext}
\begin{equation}
\label{eq:majorana-poly}
    \mathcal{U}_h^\dagger \omega \ket{\Omega_v}\bra{\Omega_v} \omega^\dagger = \sum_{p=0}^{n} \sum_{1 \leq a_1 < a_2 < \dotsb < a_{2p} \leq 2n} C_{a_1, \dots, a_{2p}} c_{a_1} c_{a_2} \dotsb c_{a_{2p}}
\end{equation}
\end{widetext}
where $2n$ is the total number of virtual Majorana modes on the link $\ell_0$, $c_i$ are the different virtual majorana modes on that link and $C_{a_1, \dots, a_{2p}}$ are scalar coefficients corresponding to each term in the expansion.
Since $\ket{\chi(\mathcal{G})}$ is a Gaussian state, one can apply equation~(17) of~\cite{bravyi_lagrangian_2005},
\begin{equation}
\label{eq:bravyi}
    \Tr \left(\rho i^p c_{a_1}c_{a_2} \dotsb c_{2p} \right)=\Pf\!\left(\tilde{\Gamma}|_{a_1,\dots, a_{2p}}\right),
\end{equation}
where $\rho$ is the density matrix of the state $\ket{\chi(\mathcal{G})}$, $1\leq a_1 < \dotsb <a_{2p}\leq 2n$, and $\tilde{\Gamma}|_{a_1,\dots a_{2p}}$ is the $2p\times 2p$ submatrix of $\tilde{\Gamma}$ (the covariance matrix of $\ket{\chi(\mathcal{G})}$) restricted to the indicated rows and columns).
Applying Eq.~\eqref{eq:bravyi} to each monomial of equation~\eqref{eq:majorana-poly} expresses the overlap as a sum of Pfaffians of submatrices of $\tilde{\Gamma}$,
\begin{widetext}
\begin{equation}
\label{eq:overlap-pfaffian}
    \bra{\chi(\mathcal{G})}\mathcal{U}_h^\dagger w \ket{\Omega_v}\bra{\Omega_v} w^\dagger\ket{\chi(\mathcal{G})}
    = \braket{\chi(\mathcal{G})}{\chi(\mathcal{G})}\sum_{p=0}^{n}\sum_{1 \leq a_1 < \dotsb < a_{2p} \leq 2n}
      C_{a_1,\dots,a_{2p}}\, i^{-p}\,\Pf\!\left(\tilde{\Gamma}|_{a_1,\dots,a_{2p}}\right),
\end{equation}
\end{widetext}
where the $p=0$ term is a scalar coefficient.

The remainder of this appendix describes how the \ggpeps package implements this reduction in practice, in four steps:
\begin{enumerate}
    \item The coefficients $C_{a_1,\dots,a_{2p}}$ are precomputed once.
    \item The covariance matrix $\tilde{\Gamma}$  of the Gaussian state $\ket{\chi(\mathcal{G})}$ is built for each gauge configuration.
    \item The required Pfaffians are computed from the covariance matrix.
    \item The results are combined into $F_E(\mathcal{G})$.

\end{enumerate}

The coefficients $C_{a_1,\dots,a_{2p}}$ depend only on the ansatz and on the single group element $h=g^{-1}g'$, not on the gauge field configuration $\mathcal{G}$ on the other links. 
Although $h$ appears to involve the gauge field $g$ on the measured link $\ell_0$, changing the integration variable in Eq.~\eqref{eq:FE} from $g'$ to $h$ turns the constraint $\bra{g'}H_E(\ell_q)\ket{g}_{\ell_0}$ into a sum of Kronecker deltas between fixed group elements and $h$ --- the same for every configuration $\mathcal{G}$.
The coefficients are therefore computed once, when the \texttt{system\_config} object is created.
It expands the operator $\mathcal{U}_h^\dagger \omega \ket{\Omega_v}\bra{\Omega_v} \omega^\dagger$ into its Majorana monomials, repeatedly simplifying the product so that every distinct monomial appears only once in the expansion. 
This simplification is achieved by bringing the modes into a fixed order and using the fact that each Majorana mode squares to the identity, after which the $i^p$ factor from Eq.~\eqref{eq:bravyi} is absorbed into the corresponding coefficient.
The scalar term, which contributes without an accompanying Pfaffian (corresponding to $p=0$ in Eq.~\eqref{eq:majorana-poly}), is kept separately. 
Iterating over the group elements $h$ that contribute, \texttt{init\_el\_energy\_terms} stores the indices for the monomials and coefficients in the arrays \texttt{idx\_vec}, \texttt{coeffs\_vec} and \texttt{constants\_vec}, indexed by group element $h$, layer, link, and monomial length for efficient vectorized evaluation.

The remaining steps all take place in the \texttt{system} class. 
The Pfaffians are evaluated on submatrices extracted from $\tilde{\Gamma}$.
Because $\ket{\chi(\mathcal{G})}$ leaves the virtual modes of the measured link $\ell_0$ uncontracted, the code treats those modes as if they were physical: the helper \texttt{extract\_mod\_covmats} reorders the system covariance matrix $\Gamma_{\textnormal{v}}$ of equation~\eqref{eq:expand_to_system} so that the modes of $\ell_0$ join the physical block, giving the modified covariance matrix $\tilde{\Gamma}_{\textnormal{v}}$.
The link covariance $\Gamma_{\textnormal{in}}$ is likewise reduced to a modified $\tilde{\Gamma}_{\textnormal{in}}$, from which the measured link $\ell_0$ has been removed, and $\tilde{\Gamma}$ is then obtained from the same contraction as equation~\eqref{eq:physical-fermions-cov}, now using $\tilde{\Gamma}_{\textnormal{in}}$ and the modified blocks $\Tilde{A}$, $\Tilde{B}$ and $\Tilde{D}$ of $\tilde{\Gamma}_{\textnormal{v}}$. The matrix inverse appearing there is cached and updated across Monte Carlo steps using the Woodbury formula rather than recomputed from scratch (as described in section~\ref{sec:mc}).
A final rotation that gauges the link $\ell_0$ using the group element $g$ completes the construction of $\tilde{\Gamma}$, which is stored in \texttt{system.covmat\_out\_mod\_vec}.

For each stored monomial, \texttt{compute\_el\_pfaffians()} then evaluates the Pfaffian of the corresponding submatrix of $\tilde{\Gamma}$, evaluating all submatrices of the same size together in a single vectorized operation. 
The coefficients are not applied at this stage.

The contributions are combined in \texttt{\_compute\_el\_energy\_op\_vec}: for each group element $h$, layer, and link it adds the constant to the coefficient-weighted sum of Pfaffians and rescales the result by the ratio of norms, where the norm of $\ket{\chi(\mathcal{G})}$ is computed from $\tilde{\Gamma}_{\textnormal{in}}$ and $\tilde{\Gamma}_{\textnormal{v}}$ through equation~\eqref{eq:norm}, giving one value per group element, layer, and link. 
The product over layers and the sum over group elements $h$ are then performed in \texttt{el\_energy\_op()}. 
Which group elements $h$ contribute, and the accompanying prefactor \texttt{el\_mult\_factor} and offset, are determined by the form of $H_E$ and are provided by the \texttt{gauge} class -- for $\mathbb{Z}_2$ a single element contributes, while for the dihedral groups only the reflections do. 
The electric energy on the whole lattice is then
\begin{equation}
\label{eq:el-energy-final}
\begin{split}
    F_E(\mathcal{G}) &= g_{E}\,\Big(\texttt{el\_offset}\cdot N_\text{links} \\
    &\quad - \texttt{el\_mult\_factor}\cdot \texttt{el\_energy\_op}\Big),
\end{split}
\end{equation}
where the offset chosen so that the energy is non-negative.

\bibliography{references.bib}
\end{document}